# Machine Learning Holography for 3D Particle Field Imaging

SIYAO SHAO,[1,2] KEVIN MALLERY,[1,2] SANTOSH KUMAR,[1,2] AND JIARONG HONG[1,2,*]

[1]Department of Mechanical Engineering, University of Minnesota, 111 Church Street SE, Minneapolis, MN 55414, USA
[2]Saint Anthony Falls Laboratory, University of Minnesota, 2 3rd Avenue SE, Minneapolis, MN 55414, USA
*jhong@umn.edu

**Abstract:** We propose a new learning-based approach for 3D particle field imaging using holography. Our approach uses a U-net architecture incorporating residual connections, Swish activation, hologram preprocessing, and transfer learning to cope with challenges arising in particle holograms where accurate measurement of individual particles is crucial. Assessments on both synthetic and experimental holograms demonstrate a significant improvement in particle extraction rate, localization accuracy and speed compared to prior methods over a wide range of particle concentrations, including highly-dense concentrations where other methods are unsuitable. Our approach can be potentially extended to other types of computational imaging tasks with similar features.

## 1. Introduction

Three-dimensional (3D) imaging for a field of particles (e.g., bubbles, droplets, aerosols, cells, etc.) has been growing exponentially in recent decades, with specific interest for measurements of size, shape, position and motion of particles for applications in flow measurements [1], environmental science [2], chemical engineering [3, 4], material sciences [5], biology [6-8], medical sciences [9-11], and others. Digital holography (DH) has recently emerged as a powerful tool for such imaging tasks with particular utility for many in situ applications [12-16] owing to its simple and compact setup. DH uses the interference between a reference wave and scattered wave from particles in a sample volume to encode the information of the particles (e.g., 3D position and size) into a hologram, and recover such information from the hologram through a digital reconstruction process. Conventional reconstruction methods convolve the holograms with diffraction kernels such as the Rayleigh-Sommerfeld and Kirchhoff-Fresnel formulas [17], and extract particle positions using image segmentation [16, 18] or focus metrics [19-21]. Image segmentation relies on prescribed intensity thresholds to distinguish the particles from the background, and its performance can deteriorate rapidly with increasing noise in the hologram which can be caused by cross interference of scattered waves from adjacent particles as the particle concentration increases [22]. Focus metric methods employ various criteria (e.g., edge sharpness, intensity distribution, etc.) to characterize the focus level of particles. These criteria are usually sensitive to detailed features of particles and the noise level in the holograms, limiting their application to low concentration particle fields with low background and cross-interference noises. To cope with these limitations, several inverse reconstruction methods such as deconvolution [23, 24] and iterative optimization [25-28] have been proposed. The deconvolution approach models the blurring observed in the 3D reconstruction as the product of convolution of the true object field with a point spread function (PSF). The PSF must be modelled based on known diffraction formulas or experimentally obtained through a hologram of a point-like object. Iterative optimization methods employ hologram formation models to minimize the difference between the observed and modeled holograms [25-28] with a set of physical constraints like sparsity and smoothness [27, 28]. However, these advanced methods are computationally intensive and require fine tuning parameters such as regularization and relaxation parameters. More importantly, the PSFs [23, 24] and hologram formation models [25-28] do not incorporate dynamic noise characteristics associated with optical distortion and particle cross-interference, which substantially hampers the performance of these methods.

Recently, machine learning, particularly using deep neural networks (DNNs), has emerged as a prevailing tool for various image analysis tasks including pattern and feature recognition in noisy images [29]. It has been successfully applied to some challenging computational imaging tasks such as extracting cell membranes from differential interference contrast images [30], improving the reconstruction of Xenopus kidney embryos from noisy confocal images [31], and recovering handwriting digits through an optically scattering slab [32]. Adoption of DNNs has drastically enhanced processing speed and yielded more accurate results than conventional inverse approaches for some applications such as imaging through a diffusive medium [29]. However, compared to other fields of computational imaging, machine learning has been under-utilized in DH [33-42]. So far, in DH, machine learning has been mostly adopted for transforming hologram reconstructions to appear as microscopic images commonly used in biological and medical applications [33-38]. Only a handful of studies have implemented machine learning for particle imaging using holography, most of which deal with single-particle holograms and using learning-based regression to extract particle depth information [39-41]. Particularly, Ren et al. [39] have demonstrated that a convolutional neural network (CNN) yields more accurate particle depth than conventional reconstruction methods and other machine learning approaches (e.g., support vector machines and k-nearest neighbor clustering). To date, only Shimobaba et al. [42] have applied machine learning for particle field reconstruction from holograms. They employed a U-net CNN architecture with an $L1$-regularized loss function and trained on synthetic holograms with particle concentration varying from $4.7\times10^{-5}$ particle/pixel (ppp) to $1.9\times10^{-4}$ ppp. Their study only demonstrated good reconstruction results for low concentration synthetic holograms with the presence of Gaussian noise and the performance of their algorithm decays rapidly with increasing particle concentrations, which is not sufficient for many practical applications. Furthermore, the $L1$ regularization employed in their approach tends to be unstable, affecting the convergence of the solution [43].

Based on the literature review and compared with other learning-based image processing tasks, we have identified three unique challenges associated with 3D particle imaging using DH. First, while the signal of an individual particle spreads over the entire hologram (the training input for machine learning), the training target usually consists of a group of sparse objects. This sparse target causes the training process to be highly unstable. Second, 3D particle field reconstruction requires very accurate measurements for each particle which differs from many conventional learning-based imaging tasks [29]. Finally, the desired metrics, recording parameters, and hologram appearance are coupled, limiting the generalizability of a model trained on a specific set of data. It is worth noting that these challenges can also appear in light field imaging [6, 44, 45], imaging through diffusive media [32], defocus imaging [46, 47] and other methods [8].

To address the above-mentioned issues, we present in this paper a specially-designed machine learning approach for 3D particle field reconstruction in DH, which can also be employed in other computational imaging tasks sharing similar traits. Section 2 describes our methodology, followed by the assessment of our method using both synthetic and experimental data in Section 3. Section 4 offers a conclusion and a discussion on the future extension of our method.

## 2. Methodology

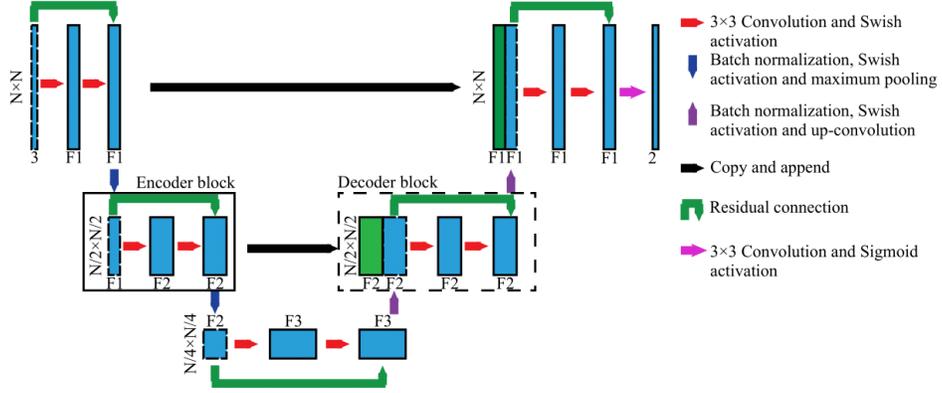

Fig. 1. The specially-designed U-net architecture for holographic reconstruction of 3D particle field.

Our machine learning approach for particle field reconstruction from a hologram uses a specially-designed U-net architecture, which takes the holograms as input and computes the 3D location of particles as output. U-net is a type of CNN developed for medical and biological image segmentation [30, 31] that has also been used in learning-based image-to-image transformations [33-37, 48] and multi-object classification from single images [49]. U-net consists of a series of encoder and decoder blocks, corresponding to solid and dashed black boxes in Fig. 1, respectively. In the encoder block, two consecutive sets of convolution layers and activation functions are used to encode local features of input images into channels. Two encoder blocks are connected by a maximum pooling layer which downsamples the feature maps in order to extract global features. The decoder block is similar but in reverse. Two consecutive convolution layers are used to decode the channels to form an image and two decoder blocks are connected by up-convolution layers to resize feature maps. Note that the output feature map of the final encoder block is connected to the first decoder block through an up-convolution layer. U-net also includes skip connections (black arrows in Fig.1) whereby the encoder output is concatenated to the same size decoder block which combines the local and global features of images for training in the deeper stage of the network. We suggest that such connections are crucial for particle field reconstruction from a hologram due to the spread of individual particle information over the entire image due to diffraction in hologram formation. In our U-net architecture, we use a U-net with 4 encoders and 3 decoders, and the number of output encoder channels are 64 and 512 for the first and last encoder, respectively. As suggested in [30], a key feature of the U-net architecture is that it can be directly applied to images of arbitrary size (regardless of the training set image size) since there are no densely connected layers.

Compared with the conventional U-net architecture, our U-net has a residual connection within each encoder and decoder block (green arrows in Fig, 1) and uses a Swish activation function for all except the last layer. The residual connection increases the training speed and reduces the likelihood of the training becoming trapped at a local minimum during stochastic gradient descent [50]. Note that within an encoder the skip connection is achieved through the connection of channels from maximum pooling to the output channels. In the decoder the residual connection uses the channels from the previous decoder block connected by an up-convolution layer. Such configuration allows the necessary shortcut connection (i.e., skipping one or two convolution layers) for a residual net [50]. Additionally, we replace the commonly-used Rectified Linear Unit (ReLU) activation function with the recently proposed Swish activation function [Eq. (1)] in our architecture [51].

$$f(x) = \frac{x}{1+e^{-x}} \qquad (1)$$

For particle holograms, the target images are usually sparse due to the small size of particle centroids which leads to the majority of values in the feature layers being equal to 0. Therefore, during the step of backpropagation in the training, the parameters within the model have a higher tendency to be 0, and subsequently untrainable since the derivative of ReLU is 0 once the weights are equal or less than 0. This problem causes substantial degradation for deep CNN models using the ReLU activation function [51]. Comparatively, the Swish function is smooth and non-monotonic near 0 which increases the number of effective parameters in the training, especially for sparse targets. However, Swish may affect the accuracy of prediction from the model due to the inclusion of negative output values. To solve this problem, we use a Sigmoid activation in the final decoder block (magenta arrow in Fig. 1) to produce results within the range from 0 to 1.

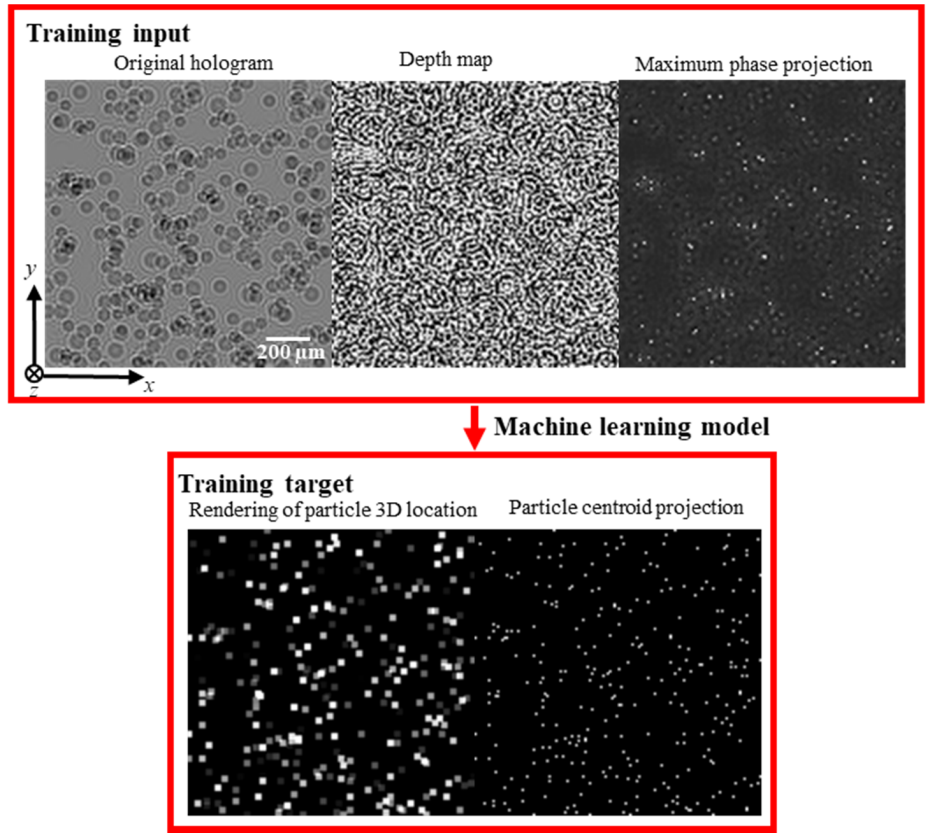

Fig. 2. A sample training input and training target consisting of 300 particles (i.e., concentration at 0.018 ppp) with a hologram size of 128 ×128 pixels. The hologram is formed with a pixel resolution of 10 μm with a laser illumination wavelength of 632 nm.

The training input consists of three channels, i.e., an original hologram, the corresponding images of pixel depth projection (i.e., depth map) and maximum phase projection (Fig. 2). The original synthetic hologram is generated following the approach in [42] and [52]. The pixel resolution is 10 μm with a laser illumination wavelength of 632 nm and an image size of 128 ×128 pixels. The particles within the holograms are randomly distributed with a distance between 1 mm and 2.28 mm to the sensor. The resolution in the longitudinal direction is also 10 μm with 128 discrete depth levels. Compared with [42], the depth map and phase projections are additional information obtained from preprocessing the holograms. As suggested by [29], preprocessing is employed to incorporate existing hologram formation knowledge into the model and reduce the need of the model to fully learn the required physics during training. Using the angular spectrum method [53], a 3D complex optical field, $u_p(x, y, z)$, is generated from the original hologram. The depth map is generated by projecting the $z$ locations where the pixels have the maximum intensity from $u_p(x, y, z)$ to $xy$ plane [Eq. (2)], and the maximum phase projection is calculated from Eq. (3).

$$z_{approx} = \arg\max_z \{u_p(x,y,z) \times \text{conj}[u_p(x,y,z)]\} \tag{2}$$

$$P(x,y) = \max_z \{\text{angle}[u_p(x,y,z)]\} \tag{3}$$

The training target consists of two output channels. The first is a grayscale channel in which the pixel intensity corresponds to the relative depth of each particle and the second is a binary image of the particle $xy$ centroids (Fig. 2). While the particles are encoded as only a single pixel in the $xy$ binary channel, doing the same for the depth-encoded grayscale channel produces a trained model which generates output with inaccurate pixel intensities and substantial background noise. To prevent this, the labeled particles in the depth-encoded grayscale target are set to a size of 3×3 pixels.

Because of the differences between the two target channels, each channel uses a different type of loss function. Specifically, a Huber loss [54] is evaluated on the output channel encoding particle depth. As shown in Eqn. 4, it uses a modified mean absolute error (MAE) of the prediction ($Y$) relative to the ground truth ($X$) as the training loss when the MAE is larger than the preset $\delta$ (0.002 for the synthetic dataset), and uses a mean squared error (MSE) when the MAE is less than $\delta$. Huber loss improves the training robustness and prediction accuracy by using MAE once the averaged pixel intensities are biased by the outliers [54]. We suggest that the parameter $\delta$ in Eq. (4) can be determined based on the measurement accuracy requirements, with a smaller $\delta$ resulting in an improved particle depth resolution. However, too small $\delta$ may lead to an unstable training process and have multiple solutions similar to using pure MAE loss [43].

$$L = \begin{cases} \frac{1}{2}\|Y-X\|_2^2 & \|Y-X\|_1 \leq \delta, \\ \delta\|Y-X\|_1 - \frac{1}{2}\delta^2 & \text{otherwise.} \end{cases} \tag{4}$$

An MSE loss regularized by the total variation (TV) of the prediction is used for the $xy$ centroid channel [Eq. (5)]. As shown in Eq. (6), TV is the sum of first-order gradients over the image of size $N_x \times N_y$.

$$L = (1-\alpha)(\|Y-X\|_2^2) + \alpha\|Y\|_{TV}^2 \tag{5}$$

$$\|Y\|_{TV} = \sum_{i=1}^{N_x}\sum_{j=1}^{N_y}\sqrt{(Y_{i,j}-Y_{i-1,j})^2 + (Y_{i,j}-Y_{i,j-1})^2} \tag{6}$$

TV regularization has previously been adopted in iterative optimization methods for hologram reconstruction [27, 28]. TV is robust to outliers in the images and causes the model to produce a smooth background in the output $xy$ centroid channel. Such regularization reduces the likelihood of background pixels having non-zero values which would result in the detection of ghost particles. The $\alpha$ in Eq. (5) is a parameter that determines the smoothness of the results. We suggest a small $\alpha$ (~0.0001) for training since TV regularization acts as a low-pass filter and too much smoothing can degrade the accuracy of the results.

The U-net architecture is implemented using Keras [55] with the TensorFlow backend. The training is conducted on a Nvidia RTX 2080Ti GPU. The Adam optimizer [56] is used with a learning rate of 0.001. Thirteen datasets, with particle concentration between $1.9\times10^{-4}$ and $6.1\times10^{-2}$ ppp, are generated to train and test the models. The highest particle concentration of the synthetic data is 305 times higher than the highest case ($1.9\times10^{-4}$ ppp) in the literature [38]. A base model is first trained on 2500 holograms with a particle concentration of $1.8\times10^{-2}$ ppp for 480 epochs (in total 13.5 hours). For each subsequent particle concentration, a dataset of 2000 holograms is trained for 120 epochs with the training initialized by the base model (2.7 hours for each case). This transfer learning approach substantially decreases the training requirement (i.e., dataset size and training time) for new hologram datasets [57]. To extract the particles from the model output, the predicted particle $xy$ centroid map is first binarized with a threshold of 0.5 to extract the $xy$ centroids of the particles. Subsequently, from the depth-encoded grayscale output, we use the intensity value of the corresponding pixels in the depth map as the particle depth.

## 3. Results

### 3.1 Assessment using synthetic holograms with constant particle concentration

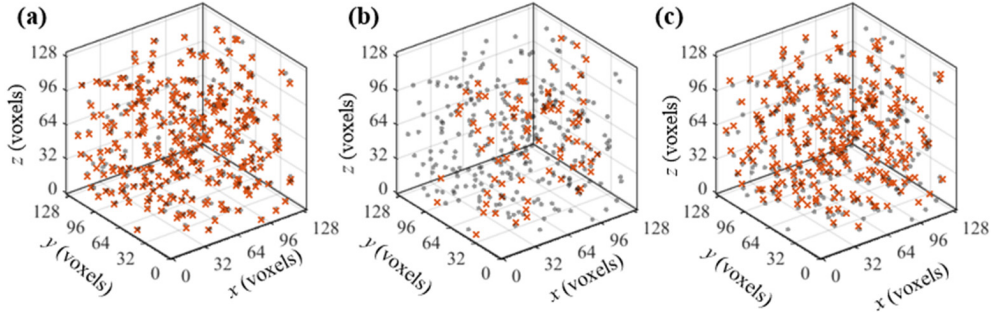

Fig. 3 Prediction results from the trained model using (a) our U-net architecture and (b) the method presented in Shimobaba et al. [42] (b) and Mallery and Hong [28] for the case of 0.018 ppp (300-particle holograms). The gray dots correspond to the particles from ground truth and red crosses are extracted particles.

We first compare the performance of our method to an implementation of the approach proposed by Shimobaba et al. [42] on a test set of 100 holograms with the same particle concentration. The training of the Shimobaba approach is conducted on a dataset of 9000 holograms with particle concentration from $1.9\times10^{-4}$ to $6.1\times10^{-2}$ ppp (same as our synthetic datasets). The comparison is first made on a 300-particle hologram (Fig. 3). The pairing of predicted particles to the ground truth follows the method presented in [28]. As Fig. 3 shows, our proposed approach yields a significantly higher number of extracted particles compared to the Simobaba approach. Our method yields an extraction rate of 98.7% with unpaired (ghost) particles accounting for about 2.3% of the total predictions. By comparison, the Shimobaba approach achieves a 40.4% extraction rate and 17.5% ghost particle rate while the state-of-the-art regularized inverse holographic volumetric reconstruction or RIHVR [28] achieves 88.4% and 2.3% for the extraction and ghost particle rates, respectively. These three methods all have a median positioning error less than 1 pixel for $x$ and $y$. The $z$ error for our method is 1.48 voxels in comparison to 5.49 voxels for Shimobaba and 3.50 voxels for RIHVR.

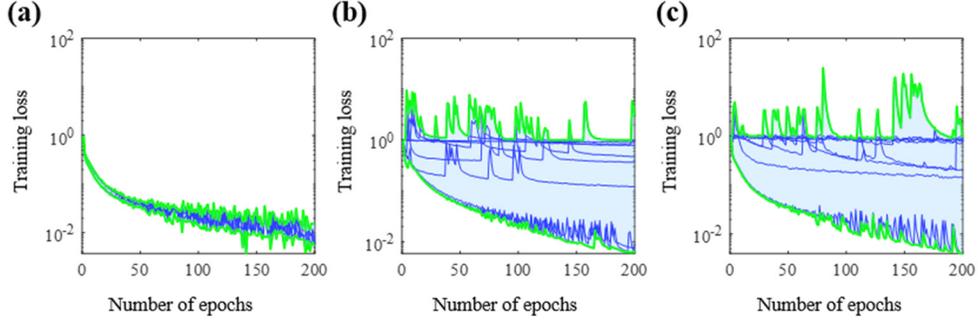

Fig.4. Demonstration of the impact of the proposed model improvements on the training process over the first 200 epochs. (a) Proposed approach, (b) using U-net architecture without residual connection and (c) using mean squared error as loss function. The loss is normalized by its initial value, and each case is randomly initialized 10 times to show the resultant instability of the training for cases (b) and (c). In the image, the green curves correspond to the maximum and minimum normalized loss at each epoch, the blue curves corresponding to each initialization, and the shaded region is the range of loss.

To analyze the impacts of the unique features in our U-net architecture on the training process, we compare the training loss decay for the first 200 epochs of different model variants on the $1.8 \times 10^{-2}$ ppp dataset. As shown in Fig. 4, compared to the proposed method, removing the residual connections or using a conventional loss function (MSE) both destabilize the training process. The removal of residual connections (Fig. 4b) leads the training process to be susceptible to local minima during training as discussed by He et al. [50]. Additionally, when using a loss function susceptible to outliers (such as MSE) without any regularization, the model is likely to produce trivial predictions in the output (i.e., spurious particles) which causes fluctuation of the loss curve even early in the training (Fig. 4c). The result is that the model does not converge to an optimal solution and produces very inaccurate pixel intensity predictions or white noise outputs for the worst scenarios.

Finally, replacement of ReLU activation functions with Swish optimizes the training process by avoiding untrainable parameters (i.e., dead neurons) [51]. From our tests, the cases without Swish can produce >6000 dead neurons in the last convolution layer at the end of the first epoch which substantially degrades the model. As a result, the training is likely to reach a plateau in the first few epochs and the resulting models generate 2D white noise images.

## 3.2 *Assessment using synthetic holograms with variable particle concentration*

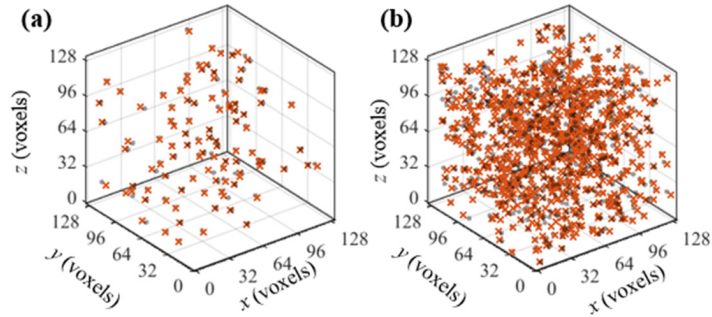

Fig. 5. Comparison of prediction results with a 100-particle hologram (a) and a 1000-particle hologram (b). The gray dots correspond to ground truth and red crosses are extracted particles.

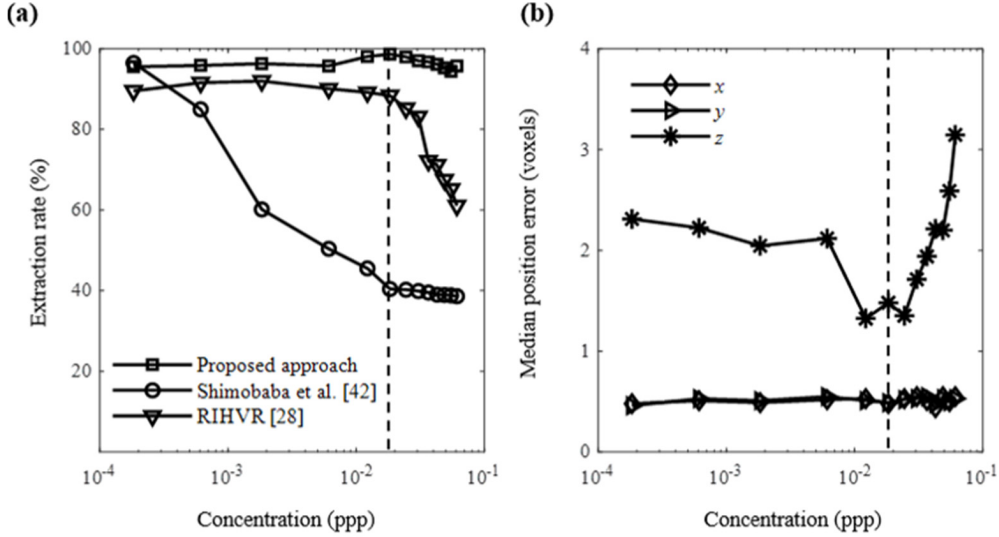

Fig. 6. (a) Extraction rate under different particle concentrations of the proposed method and compared with the case of Shimobaba et al. [42] and RIHVR [28] and (b) Median position error of extracted particles for the proposed method under different particle concentrations. Note that the dashed lines correspond to the particle concentration of the base model ($1.8\times10^{-2}$ ppp).

The particle extraction rate and positioning accuracy using the proposed transfer learning approach (see Section 2) are assessed for variable particle concentrations from $1.9\times10^{-4}$ ppp to $6.1\times10^{-2}$ ppp. In Fig. 5, we present a case of 100-particle hologram ($6.1\times10^{-3}$ ppp) and a case of 1000-particle hologram ($6.1\times10^{-2}$ ppp). The lowest concentration instance shown here has an extraction rate of 98.0% while the highest concentration case has a 95.0% extraction rate. From an assessment of 100 holograms for each particle concentration, the extraction rate is above 94.4% over the range for $1.9\times10^{-4}$ ppp to $6.1\times10^{-2}$ ppp (Fig. 6a) with a median particle positioning accuracy less than one voxel for $x$ and $y$ and less than 3.2 voxels for $z$ for all concentrations (Fig. 6b). For all cases, the ghost particle ratio of the model prediction is less than 10%. Using the same test data, Shimobaba et al. [42] yields an extraction rate lower than 60% for the particle concentrations higher than $1.8\times10^{-3}$ ppp and RIHVR [28] shows a substantial drop in particle extraction rate starting at a concentration of $3.0\times10^{-2}$ ppp. Our machine learning approach shows no such drop within the studied range (Fig. 6a). We suggest that our pre-processing and transfer learning approaches significantly reduce the training requirements to yield a high extraction rate and positioning accuracy for new holograms especially for high concentration cases. The increased extraction rate for high particle concentration holograms potentially enables improved spatial resolution tracer-based flow diagnostic techniques such as particle image velocimetry (PIV) and particle tracking velocimetry (PTV) [1].

### 3.3 *Assessment using experimental data*

The proposed method is assessed using experimental holograms of fluorescently labeled particles embedded in a solid gel. We use 2 µm fluorescent particles (ThermoScientific) at a concentration of ~5000 particles/mm$^3$ ($2.0\times10^{-2}$ ppp) dispersed in a water-based gelatin placed between glass slides. The experimental holograms (Fig. 7a) are recorded on a Nikon Ti-Eclipse inverted microscope using a 10X Nikon objective lens and an Andor Zyla 5.5 sCMOS (0.65 µm/px) with a collimated 660 nm diode laser illumination (QPhotonics; QLD-660-10S). The microscope can record multiple holograms of 2432×2048 pixels at distances spanning 0-200 µm below the volume all of which are used to calculate an ensemble averaged background for image enhancement. The ground truth measurement (Fig. 7b) is obtained by scanning the sample at the same location using the epi-fluorescence mode of the microscope at a step size of 2.5 μm over the entire depth at the same image size and resolution. The particle positions for the ground truth measurement are then obtained through manual thresholding and intensity weighted centroid calculation. The training dataset consists of 7500 randomly cropped 128×128 pixel tiles from the enhanced hologram and their corresponding particle locations from the 3D fluorescence scan volume. Here the target images are saved with 16-bit precision to encode a large number of reconstruction planes. The Huber loss $\delta$ in Eqn. 4 is set as 0.001 which is lower than the synthetic cases since the pixel intensity in the labels encode higher resolution in the depth ($y$ direction in the experimental case). We use the same method as the synthetic case to pair the predicted particles to the ground truth. As shown in Fig. 7(c), despite the noisy input (Fig. 7a), the predicted results from the trained model match well to the ground truth. The test of 100 randomly cropped 128×128 pixel tiles from a validation hologram imaging a different region of this sample yields a 90% extraction rate with a positioning error less than 1 voxel in the $x$ and $y$ directions and 5.24 voxels in the $z$ direction. The training of this dataset using Shimobaba approach [42] cannot converge to a low loss value and yields a model generating strong background noise. It is difficult to apply RIHVR to this case because the background removal specified in [28] depends on the motion of particles to produce an accurate time-averaged background image. As such, RIHVR has a very high false detection rate for this case.

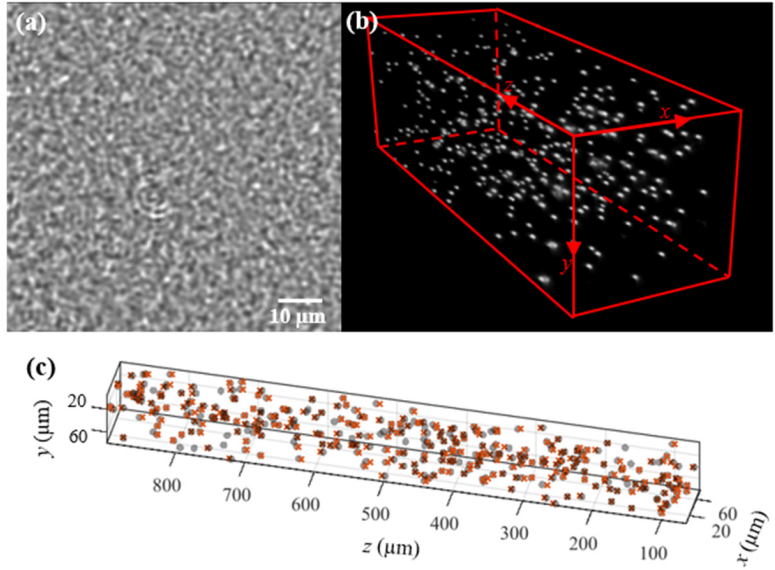

Fig. 7. (a) A 128×128-pixel enhanced hologram from the experimental data and corresponding volumetric image through the stacking of fluorescent bright field scanning of the same sample for determining the ground truth (b). (c). Prediction results in comparison with the ground truth. The gray dots correspond to ground truth and red crosses are extracted particles.

## 4. Summary and discussion

In the present paper, we introduce a new learning-based approach for 3D particle field reconstruction from holograms. For holograms, the information of individual particles is usually spread over the whole image (training input) while the training target consists of a sparse particle field where accurate measurement of each particle is crucial. To handle these traits, our specially-designed U-net architecture has three input channels (original holograms, depth, and maximum phase projection maps) and two output channels (depth and centroid maps of reconstructed particles). The 2D depth and maximum phase projection map channels use the angular spectrum method from conventional holographic reconstruction to incorporate hologram formation knowledge into the U-net training and reduce the need to fully learn the required physics. In addition, our architecture employs residual connections and the Swish activation function to reduce the likelihood of the training becoming trapped in local minima or producing a large number of dead neurons. We use two types of loss functions, Huber loss and TV-regularized MSE loss, for the output channels of particle depth and particle centroids, respectively, in order to improve the prediction accuracy, produce a smooth background, and reduce ghost particles. Lastly, a transfer learning approach is adopted to reduce the training requirements for new hologram datasets. Through an assessment of synthetic holograms, our approach is suitable for processing much denser particle concentrations than prior approaches, with a 94% extraction rate at a concentration ($6.1 \times 10^{-2}$ ppp) 305 times higher than previously demonstrated with a machine learning approach [42] and 4 times higher than the 90% extraction limit for a state-of-the-art analytical method [28]. This improvement to the maximum concentration comes while also achieving improved positioning accuracy (error of <3.2 voxels). Validating the proposed method on experimental holograms with a concentration of 0.020 ppp results in an extraction rate over 90% with a positioning error of 5.24 voxels for the depth measurement. These assessments demonstrate the unique power of machine learning for particle hologram reconstruction with a broad range of particle concentrations. Finally, our learning-based hologram reconstruction is more than 30 times faster than the analytical RIHVR method, even though minimal effort has been undertaken to optimize the speed of the current approach. We suggest that the proposed method can be generalized for sparse field imaging tasks such as imaging individual brain neuron activities, particle localization with synthetic aperture or defocusing imaging, and imaging through diffusive media.

While our machine learning approach has equal or superior performance compared to the state-of-the-art conventional hologram reconstruction method, there remains room for improvement. First, new ground truth data and additional training are usually required to achieve the optimal performance of our approach for new experimental datasets. Nevertheless, transfer learning adopted in our approach can substantially reduce the training time for new datasets to reach an accurate 3D particle field reconstruction (for example, 2.7 hours training on a new dataset compared to 13.5 hours without transfer learning). For holograms collected with significantly different recording settings, the ground truth can be collected using experiments or through high-fidelity conventional reconstruction approaches such as RIHVR [28]. Moreover, ongoing work is aiming to synthesize holograms with sufficient fidelity to train our U-net architecture suitable for processing experimental images. Such an approach can substantially reduce the cost of collecting ground truth measurements and has been proven effective for image classification tasks [58] and 2D shadow image particle segmentation [59]. Additionally, although our learning-based approach has achieved significant speed improvement in comparison to conventional reconstruction methods, more than 90% of our total processing time is consumed in pre- and post-processing steps. The processing speed of such steps can be readily enhanced through GPU processing and a more streamlined pipeline to attain real-time/onboard processing capacity for various applications.

## Funding

Office of Naval Research (ONR) (N000141612755).

## Acknowledgements

The authors would like to thank Prof. Xiang Cheng for access to the microscope used to generate the experimental training data.

## Disclosures

The authors declare no conflicts of interest.